%% file: O3-j0537-paper.tex
\documentclass[fleqn,usenatbib,useAMS,twocolumn]{aastex62}
\usepackage[utf8]{inputenc}

\usepackage{color}
\usepackage{graphicx}
\usepackage{amsmath}
\usepackage{amssymb}
\usepackage{bm}
\usepackage{acronym}
\usepackage{ifthen}
\usepackage{blindtext}
\usepackage[normalem]{ulem}
\usepackage{hyperref}
\usepackage{etoolbox}
\usepackage{fancyhdr}
\usepackage{xspace}
\usepackage{textcomp}
\usepackage{multirow}
\usepackage[caption=false]{subfig}
\usepackage{lineno}
\usepackage{tabularx}

\def\gw{gravitational wave\xspace}
\def\gwh{gravitational-wave\xspace}
\def\gws{gravitational waves\xspace}

\newacro{S/N}{signal-to-noise ratio}

\newacro{PN}{post-Newtonian}

\newacro{O3}{the third observing run}

\newacro{O2}{the second observing run}

\newacro{MSP}{millisecond pulsar}

\newacro{CW}{\emph{Continuous Wave}}

\def\NICER{{\it NICER}\xspace}
\def\RXTE{{\it RXTE}\xspace}

\def\psrname{PSR~J0537\textminus6910\xspace}
\def\nus{f_{\rm rot}}
\def\nudot{\dot{f}_{\rm rot}}
\def\nuddot{\ddot{f}_{\rm rot}}
\def\nig{n_{\rm ig}}

\newtoggle{fullauthorlist}
\toggletrue{fullauthorlist}
\newtoggle{endauthorlist}
\toggletrue{endauthorlist}

\newcommand\scinum[2]{\ensuremath{#1\!\times\!10^{#2}}}

\shorttitle{\gws from PSR~J0537\textminus6910}
\shortauthors{Abbott et al.}

\begin{document}
\title{Diving below the spin-down limit: Constraints on gravitational waves from the energetic young pulsar PSR~J0537\textminus6910}

\iftoggle{endauthorlist}{
  %
  %
  \let\mymaketitle\maketitle
  \let\myauthor\author
  \let\myaffiliation\affiliation
  \author{The LIGO Scientific Collaboration}
  \author{The Virgo Collaboration}
  \author{The KAGRA Collaboration}
  \email{Full author list given at the end of the article.}
\noaffiliation
}{
\iftoggle{fullauthorlist}{
  \input{KAGRA-LSC-Virgo-Authors-Aug-2020-aas.tex}
  \input{authorlist_nicer.tex}
}
{
  \author{The LIGO Scientific Collaboration}
  \affiliation{LSC}
  \author{The Virgo Collaboration}
  \affiliation{Virgo}
  \author{The KAGRA Collaboration}
  \affiliation{KAGRA}
  
  \author{D. Antonopoulou}
  \affiliation{Nicolaus Copernicus Astronomical Center, Polish Academy of Sciences, ul.\ Bartycka 18, 00-716 Warsaw, Poland}
  \author{Z. Arzoumanian}
  \affiliation{X-Ray Astrophysics Laboratory, NASA Goddard Space Flight Center, Greenbelt, MD 20771, USA}
  \author{T. Enoto}
  \affiliation{Extreme Natural Phenomena RIKEN Hakubi Research Team, RIKEN Cluster for Pioneering Research, 2-1 Hirasawa, Wako, Saitama 351-0198, Japan}
  \author{C.~M. Espinoza}
  \affiliation{Departamento de F\'isica, Universidad de Santiago de Chile, Avenida Ecuador 3493, 9170124 Estaci\'on Central, Santiago, Chile}
  \author{S. Guillot}
  \affiliation{IRAP, CNRS, 9 avenue du Colonel Roche, BP 44346, F-31028 Toulouse Cedex 4, France}
  \affiliation{Universit\'e de Toulouse, CNES, UPS-OMP, F-31028 Toulouse, France}
}

}
\date{\today}

\begin{abstract}
We present a search for quasi-monochromatic gravitational-wave signals from the young, energetic X-ray
pulsar PSR~J0537\textminus6910 using data from the second and third observing runs of LIGO and Virgo. The search is enabled by a contemporaneous timing ephemeris obtained using \NICER data.  The \NICER ephemeris has also been extended through October 2020 and includes three new glitches.
PSR~J0537\textminus6910 has the largest spin-down
luminosity of any pulsar and exhibits frequent and strong glitches. Analyses of its
long-term and inter-glitch braking indices provide intriguing evidence that its spin-down
energy budget may include gravitational-wave emission from a time-varying mass quadrupole moment. 
Its 62\,Hz rotation frequency also puts its possible
\gwh emission in the most sensitive band of the LIGO/Virgo detectors. 
Motivated by these considerations, we search for \gwh emission at both once and twice the rotation frequency from PSR J0537\textminus6910. We find no signal, however, and report upper limits. 
Assuming a rigidly rotating triaxial star, our constraints reach below the gravitational-wave spin-down limit for this star for the first time by more than a factor of two and limit \gws from the
$l=m=2$ mode to account for less than 14\% of the spin-down energy budget. The fiducial equatorial ellipticity is
constrained to less than about \scinum{3}{-5}, which is the third best constraint for any young pulsar. 
\end{abstract}

\keywords{gravitational waves --- pulsars: general --- pulsars: individual (\psrname) --- stars: neutron}

\section{Introduction}\label{sec:intro}

The young (1--5~kyr) energetic pulsar \psrname \citep{wangetal98,chenetal06} resides in the Large Magellanic Cloud at a distance of 49.6~kpc \citep{pietrzynskietal19}. Its pulsations are only detectable at X-ray energies, and the pulsar was first observed
by \citet{1998ApJ...499L.179M} using the {\it Rossi X-ray Timing Explorer} (\RXTE) during searches
for pulsations from the remnant of SN1987A. Further observations with \RXTE, prior to its decommissioning in
early 2012, revealed that \psrname often undergoes sudden changes in rotation frequency, i.e., glitches, at a rate of more than three per year, and exhibits interesting inter-glitch behavior \citep{2004ApJ...603..682M, 2006ApJ...652.1531M,anderssonetal18, antonopoulouetal18,2018ApJ...852..123F}.
Observations of the pulsar resumed from 2017-2020 using the {\it Neutron star Interior 
Composition Explorer} (\NICER) on board the International Space Station \citep{2012SPIE.8443E..13G}, which revealed more glitches and a continuation of the timing behavior seen with \RXTE \citep{hoetal20}.

\psrname is a particularly intriguing potential \gwh
source.  It is the fastest-spinning known young pulsar (with rotation frequency
$\nus=62\mbox{ Hz}$), which places its \gwh frequency $f$ (e.g., at twice $\nus$; see Section~\ref{sec:model}) in the most sensitive band of ground-based \gwh detectors. \psrname also has the highest spin-down luminosity ($\dot{E}=4.9\times 10^{38}\mbox{ erg s$^{-1}$}$) among
the $\sim$2900 known pulsars in the ATNF Pulsar Catalogue \citep{2005AJ....129.1993M}.  Its spin-down behavior appears to be driven by a process
other than pure electromagnetic dipole radiation loss (at constant stellar
magnetic field and moment of inertia). Specifically, its (long-term) braking index
$n\equiv\nus\nuddot/\nudot^2=-1.25\pm0.01$, as measured over more than 20~yr \citep{hoetal20},
indicates an accelerating spin-down rate and significantly deviates from the measured values of most pulsars, $n=3$, that imply dipole radiation \citep{shapiroteukolsky83}.

\begin{figure}[htb]
\begin{center}
\includegraphics[width=\columnwidth]{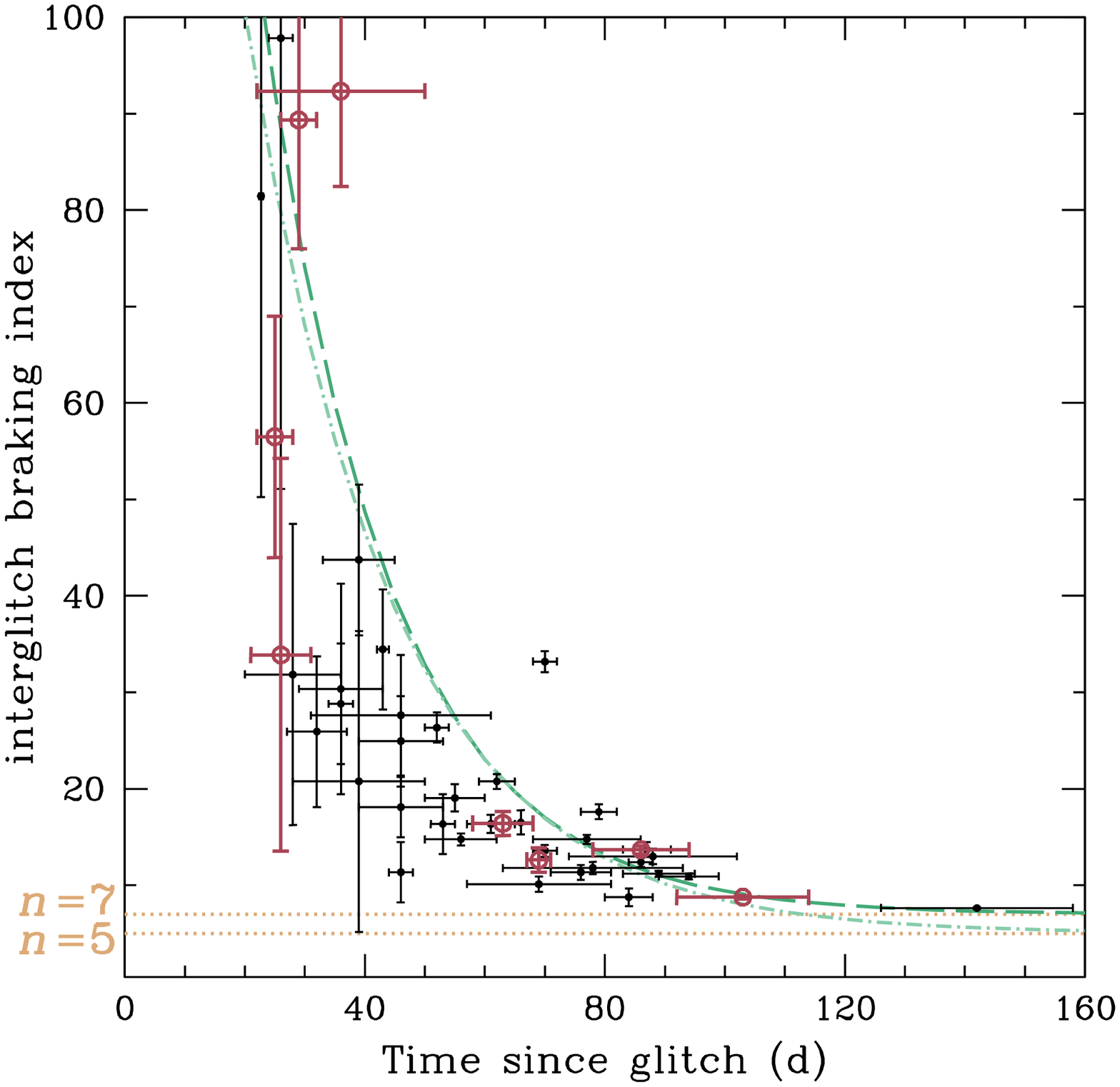}
\caption{
Interglitch braking index $\nig$ calculated from the spin parameters of each segment
between glitches as a function of time since the last glitch.
Large and small circles denote \NICER\ and \RXTE\ values, respectively,
with the former from Tables~\ref{tab:model} and \ref{tab:glitch} and from \citet{hoetal20} and latter from \citet{antonopoulouetal18}.
Errors in $\nig$ are 1$\sigma$ uncertainty.
Orange horizontal dotted lines indicate braking index $n=5$ and 7, which are
expected for pulsar spin-down by \gwh emission due to an ellipticity and
r-mode oscillation, respectively.
Green dot-dashed and dashed lines indicate exponential decay to $n=5$
with best-fit time-scale of 24~d and to $n=7$ with best-fit time-scale
of 21~d, respectively.
}
\label{fig:tnig}
\end{center}
\end{figure}

More importantly,
observations of \psrname show the pulsar's (short-term) interglitch braking index $\nig$, as measured during intervals between
$\sim 50$ glitches, has values typically $> 10$, and approaches an asymptotic
value of $\lesssim 7$ at long times after a glitch, i.e., when the effects
of a preceding glitch are diminished (see Figure~\ref{fig:tnig}; see also \citealt{anderssonetal18}).
It is this behavior that provides tantalizing suggestions that \psrname could be
losing some of its rotational energy to \gwh emission.
In particular, a slightly deformed pulsar can emit \gws that results in $n=5$, and a r-mode fluid oscillation in a pulsar can emit \gws that results in $n=7$ (see, e.g., \citealt{riles17,anderssonetal18,2018ASSL..457..673G,gao20}).

In this work, we search for mass quadrupolar \gwh emission from \psrname that follows the same phase as that of the pulsar's rotation. Previously, data from initial LIGO's fifth and sixth science runs (S5 and S6) and Virgo's second and fourth science runs (VSR2 and VSR4), in conjunction with \RXTE timing measurements, were used to set limits on \gwh emission by \psrname that closely approached the spin-down limit \citep{2010ApJ...713..671A,2014ApJ...785..119A}. Here, we analyze data from the second and third observing runs (O2 and O3) of LIGO and Virgo, tracking the rotation phase with the contemporaneous \NICER timing ephemeris.
In doing so, we also provide an updated ephemeris that includes the latest six months of \NICER observations of \psrname.

Investigations of r-mode \gwh emission ($n=7$) are not presented here; such searches are more technically challenging and require different methods that search over a range of frequencies (see, e.g.,  \citealt{mytidis2015constraining,mytidis2019sensitivity,O2Narrowband,fesikpapa20,fesikpapa20b}) due to uncertainty in \gwh frequency for a given rotation frequency \citep{anderssonetal14,idrisyetal15,carideetal19}. 
Nevertheless, we are able to reach below the spin-down limit of \psrname for the first time, which means that the minimum amplitude we could detect in our analysis is lower than the one obtained by assuming all of the pulsar's rotational energy loss is converted to \gws (see Section~\ref{sec:model}). In other words, we can now obtain physically meaningful constraints.

\section{Search method}
\subsection{Model of \gwh emission} \label{sec:model}

The first model considered here allows for \gwh emission at once and twice the spin frequency simultaneously, which has been searched for previously \citep{2015MNRAS.453.4399P,2017ApJ...839...12A, 2019ApJ...879...10A,abbott2020gravitational}, and can result from a triaxial star spinning about an axis that is not its principal axis \citep{2010MNRAS.402.2503J,2015MNRAS.453...53J}.
The amplitudes of each harmonic at once and twice the spin frequency of the star, denoted $h_{21}(t)$ and $h_{22}(t)$, respectively, can be written as
\begin{eqnarray}\label{eq:model}
\nonumber
h_{21} &=& - \frac{C_{21}}{2} \Big\{ F^D_+(\alpha, \delta, \psi; t) \sin\iota \cos\iota \cos{\left[\Phi(t)+ \Phi_{21}^C\right]}\\
 && \qquad +F^D_\times(\alpha, \delta, \psi; t)\sin\iota  \sin{\left[\Phi(t)+ \Phi_{21}^C\right]} \Big\}, \label{eq:model1} \\
\nonumber
h_{22} &=& - C_{22} \Big\{ F^D_+(\alpha, \delta, \psi; t) (1 + \cos^2\iota) \cos{\left[2\Phi(t)+ \Phi_{22}^C\right]}   \\
&& \qquad +2 F^D_\times(\alpha, \delta, \psi; t) \cos\iota  \sin{\left[2\Phi(t)+ \Phi_{22}^C\right]} \Big\}. \label{eq:model2}
\end{eqnarray}
Here, $C_{21}$ and $C_{22}$ are dimensionless constant component amplitudes, and $\Phi_{21}^C$ and $\Phi_{22}^C$ are phase angles.  $F^D_+$ and $F^D_\times$ are antenna or beam functions and describe how the two polarization components of the signal are projected onto the detector \citep[see, e.g.,][]{1998PhRvD..58f3001J}. The angles $(\alpha, \delta)$ are the right ascension and declination of the source, while the angles $(\iota, \psi)$ specify the orientation of the star's spin axis relative to the observer.  $\Phi(t)$ is the rotational phase of the source.

The second model is a special case of the first model and is used for \gwh emission at only twice the rotational frequency ($C_{21}=0$), implying a triaxial star that is spinning about a principal axis, such as its z-axis. In this case, it is simpler to write the \gwh amplitude in terms of the dimensionless value $h_0$, which requires substituting $C_{22} = -h_0 / 2$ in equation~(\ref{eq:model2}) \citep{2019ApJ...879...10A}. The sign change simply maintains consistency with the model from \citet{1998PhRvD..58f3001J}. The cause of such \gwh emission is a deviation from axial symmetry, which can be written in terms of a dimensionless equatorial ellipticity $\varepsilon$, defined in terms of the star's principal moments of inertia $(I_{xx}, I_{yy}, I_{zz})$:

\begin{equation}\label{eq:ellipticity}
    \varepsilon \equiv \frac{|I_{xx} - I_{yy}|}{I_{zz}}.
\end{equation}
The \gwh amplitude is directly proportional to the ellipticity:
\begin{equation}\label{eq:h0}
    h_0 = \frac{16\pi^2 G}{c^4} \frac{I_{zz} \varepsilon \nus^2 }{d},
\end{equation}
where $d$ is the star's distance from the Earth. When setting upper limits, we use a fiducial value for the z-component of the moment of inertia, \emph{i.e.}, $I_{zz}^{\rm fid} = 10^{38}$\,kg\,m$^2$.
The combination of the ellipticity and fiducial moment of inertia can be cast in terms of the mass quadrupole moment of the $l=m=2$ mode of the star via $Q_{22} = \sqrt{15/8\pi}I_{zz} \varepsilon$ \citep{2005PhRvL..95u1101O}.
The \gwh amplitude $h_0$ can be compared to the spin-down limit amplitude $h_0^{\rm sd}$, which is the \gwh amplitude produced assuming that all of the rotational energy lost by the pulsar is converted into \gws:

\begin{equation}\label{eq:h0sd}
    h_{0}^{\rm sd}=\frac{1}{d}\left(\frac{5GI_{zz}}{2c^3}\frac{|\dot{f}_{\rm rot}|}{f_{\rm rot}}\right)^{1/2}.
\end{equation}
Our results for the single harmonic case are quoted in terms of $h_{0}^{\rm sd}$.

\NICER observations of \psrname allow for the ephemeris of the pulsar to be determined, which means
we know the expected signal frequency and its evolution.
With this information, we can perform a targeted search for \gws from this pulsar based on the two signal models  discussed, with the phase tracking that of the pulsar rotation.

\subsection{\NICER data} \label{sec:nicer}

\begin{figure}
\begin{center}
\includegraphics[width=\columnwidth]{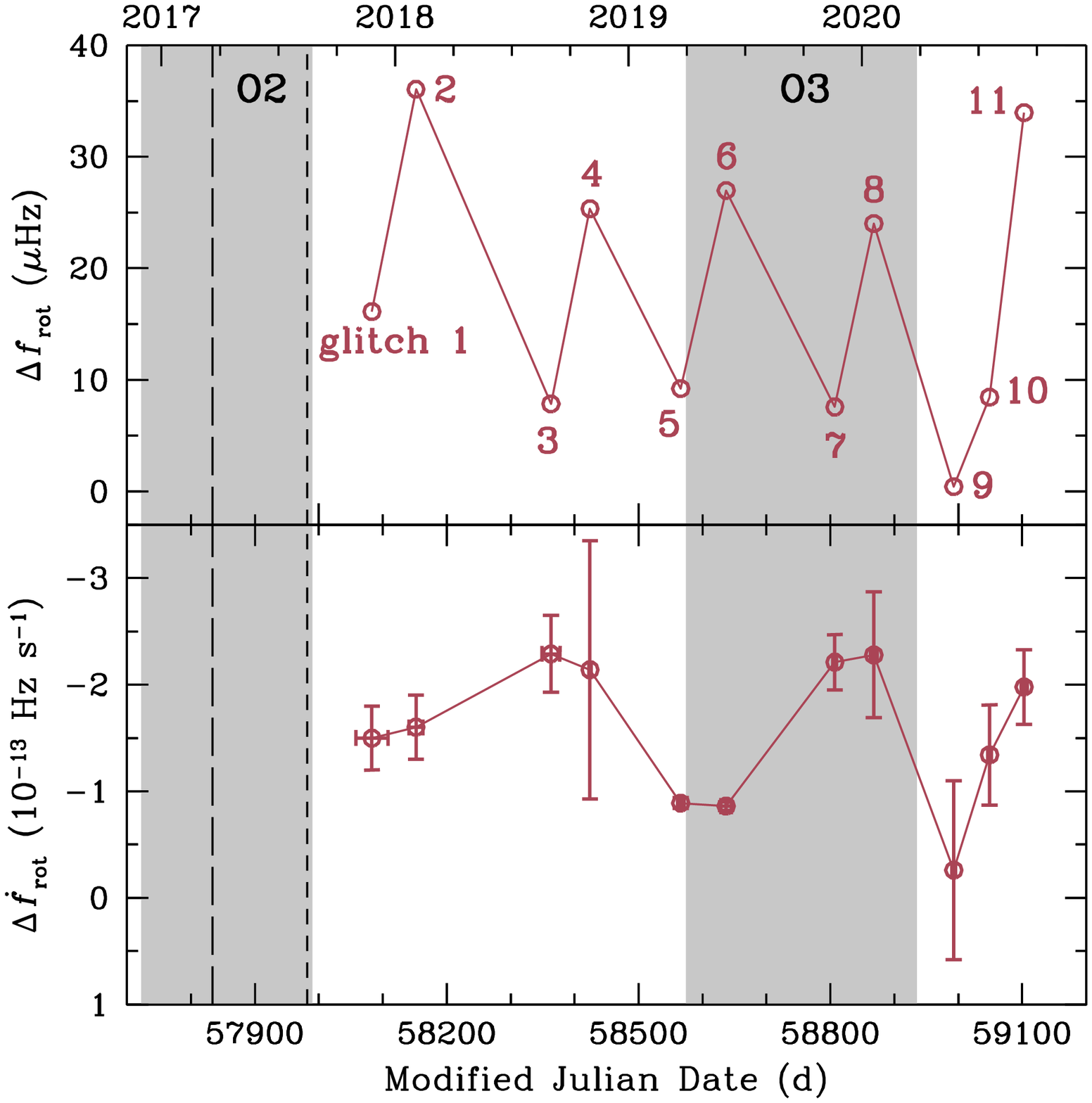}
\caption{
Glitch $\Delta\nus$ (top) and $\Delta\nudot$ (bottom) as functions of time.
Glitch numbers and values from Table~\ref{tab:glitch} and \citet{hoetal20}.
Errors in $\Delta\nudot$ are 1$\sigma$ uncertainty, while errors in $\Delta\nus$ are
not shown because they are generally smaller than the symbols.
Shaded regions denote second observing run (O2) and third observing run (O3) of LIGO/Virgo.  Vertical long and short-dashed lines indicate two possible start dates of O2 data used in the present work (see Section~\ref{sec:gwdata}).
}
\label{fig:glitch}
\end{center}
\end{figure}

In \citet{hoetal20}, timing analysis is performed on \NICER data of
\psrname from 2017 August 17 to 2020 April 25, with eight glitches detected
during this timespan and the last three glitches during O3.
Here we present an update and results on timing analysis since the work
of \citet{hoetal20}.  In particular, data from 2020 May 12 to October 29
is analyzed using the methodology as described in \citet{hoetal20}.
Our analysis reveals continuing accelerated spin-down (see
Table~\ref{tab:model}) and
three subsequent glitches (see Table~\ref{tab:glitch} and
Figure~\ref{fig:glitch}), including the smallest glitch of \psrname
yet detected using \NICER.
Note that the timing model of segment 8 uses three additional subsequent
times-of-arrival (TOAs) beyond those in Table~1 of \cite{hoetal20}
and, as a result, the epoch and other parameters of the model differ;
e.g., {segment 8 is associated with the data point at 63~d and $\nig=16$ in Figure~\ref{fig:tnig} compared to 50~d and $\nig=22$ in} Figure~6 of
\citet{hoetal20}.
Meanwhile, the relatively short timespan of segment 9 means
the timing model for this segment is not able to constrain 
$\nuddot$.
For the most recent glitch 11, its magnitude is large ($\Delta\nus=33.9\mbox{ $\mu$Hz}$), which suggests the time to the next glitch will be long ($\sim 200\pm 20\mbox{ d}$; \citealt{hoetal20}).  If the interglitch period is indeed long, then \NICER measurements could eventually yield $\nig\lesssim 7$ for segment 11, which would lend further support for \gwh emission (see Section~\ref{sec:intro} and Figure~\ref{fig:tnig}).

\begin{deluxetable*}{ccccccccccc}[htb]
\tablecaption{Timing model parameters for segments between epochs of new glitches of \psrname. Columns from left to right are segment number, timing model epoch, segment start and end dates, number of times-of-arrival, rotation frequency and its first two time derivatives, interglitch braking index, and timing model residual and goodness-of-fit measure.  Number in parentheses is 1$\sigma$ uncertainty in last digit. Segments 1--7 are presented in \cite{hoetal20}.
\label{tab:model}}
\tablewidth{0pt}
\tablehead{
\colhead{Segment} & \colhead{Epoch} & \colhead{Start} & \colhead{End} & \colhead{TOAs} & \colhead{$\nus$} & \colhead{$\nudot$} & \colhead{$\nuddot$} & \colhead{$\nig$} & \colhead{Residual RMS} & \colhead{$\chi^2/\mbox{dof}$} \\
 & \colhead{(MJD)} & \colhead{(MJD)} & \colhead{(MJD)} & & \colhead{(Hz)} & \colhead{($10^{-10}$ Hz s$^{-1}$)} & \colhead{($10^{-20}$ Hz s$^{-2}$)} & & \colhead{($\mu$s)} &
}
\startdata
 8 & 58931 & 58871.5 & 58991.2 & 17 & 61.908808739(3) & $-1.997535(7)$ & 1.06(8) & 16(1) & 173.7 &  9.9 \\
 9 & 59020 & 58995.6 & 59046.3 & 11 & 61.907273376(2) & $-1.99699(4)$ & [1]\tablenotemark{a} & --- & 147.8 &  6.7 \\
10 & 59074 & 59050.4 & 59098.7 &  10 & 61.906349948(5) & $-1.99762(2)$ &  3.6(8) & 56(13) &  60.9 &  1.5 \\
11 & 59129 & 59108.7 & 59150.7 &  11 & 61.905434556(6) & $-1.99809(3)$ & 2.2(13) & 34(20) &  72.3 &  2.1 \\
\enddata
\tablenotetext{a}{ $\nuddot$ is fixed at $10^{-20}\mbox{ Hz s$^{-2}$}$.}
\end{deluxetable*}

\begin{deluxetable*}{cccccc}[htb]
\tablecaption{Parameters of new glitches of \psrname. Columns from left to right are glitch number and epoch, change in rotation phase and changes in rotation frequency and its first two time derivatives at each glitch.  Number in parentheses is 1$\sigma$ uncertainty in last digit. Glitches 1--7 are presented in \cite{hoetal20}.
\label{tab:glitch}}
\tablewidth{0pt}
\tablehead{
\colhead{Glitch} & \colhead{Glitch epoch} & \colhead{$\Delta\phi$} & \colhead{$\Delta\nus$} & \colhead{$\Delta\nudot$} & \colhead{$\Delta\nuddot$} \\
 & \colhead{(MJD)} & \colhead{(cycle)} & \colhead{($\mu$Hz)} & \colhead{($10^{-13}$ Hz s$^{-1}$)} & \colhead{($10^{-20}$ Hz s$^{-2}$)}
}
\startdata
 8 & 58868(5) & $ 0.08(12)$ & 24.0(1)  & $-2.3(6)$  & $-5(1)$ \\
 9 & 58993(3) & $ 0.06(12)$ &  0.4(1)  & $-0.3(8)$  & --- \\
10 & 59049(3) & $-0.22(2)$  &  8.46(3) & $-1.3(5)$  & --- \\
11 & 59103(5) & $0.42(2)$  &  33.958(7)  & $-2.0(3)$  & --- \\
\enddata
\end{deluxetable*}

The \gwh search performed here uses the timing model of \citet{hoetal20}. The differences between the model of \citet{hoetal20} and the model presented here are well within the former's uncertainties, and thus use of the latter would not yield significantly different results.

\subsection{LIGO and Virgo data}\label{sec:gwdata}
We use a combination of data from the second and third observing runs
of the Advanced LIGO \citep{2015CQGra..32g4001L} and Virgo \citep{2015CQGra..32b4001A} gravitational wave
detectors. During O2, LIGO Livingston (L1) and LIGO Hanford (H1) took data from
2016 November 30 to 2017 August 25 and had duty factors of $\sim 57\%$ and $\sim 59\%$, respectively (including commissioning breaks), while Virgo took data from 2017 August 1 to 2017 August 25 with a duty factor of $\sim 85\%$.
As noted in Section~\ref{sec:nicer}, \NICER data start on 2017 August 17,
and thus one set of searches we undertake uses only about six days of O2 data overlapping with the NICER data in addition to the O3 data {(explicitly 5.3, 5.5 and 6.0 d of data for H1, L1 and V1, respectively)}.  Alternatively, we can consider a more optimistic and much longer time-series of O2 data by taking advantage of the correlation between glitch size and time-to-next-glitch seen for \psrname \citep{2006ApJ...652.1531M,antonopoulouetal18,2018ApJ...852..123F,hoetal20}.  Assuming a (unobserved) glitch occurred on 2017 March 22 with the same size as the largest \NICER glitch (i.e., glitch 2 with $\Delta\nus=36\mbox{ $\mu$Hz}$), we would expect a subsequent glitch 224 d later (at 68\% confidence) on 2017 November 1, which is the earliest estimated date at which glitch 1 occurred (see Figure~\ref{fig:glitch} and \citealt{hoetal20}). Thus 2017 March 22 to November 1 is the longest period over which we would expect \psrname to not have undergone a glitch and the \NICER ephemeris to be valid.
O3 lasted from 2019 April 1 to 2020 March 27, with a one-month pause in data collection in October 2019. The three detectors' datasets H1, L1, and V1 had duty factors of $\sim {72}\%$, $\sim {69}\%$, and $\sim {69}\%$, {or 259, 248 and 248 d of data,} respectively during O3.

In the case of a detection, calibration uncertainties limit our ability to provide robust estimates of the amplitude of the \gwh signal and corresponding ellipticity \citep{Abbott:2017pqa}. Even without a detection, these uncertainties affect the estimated instrument sensitivity and inferred upper limits. The uncertainties vary over the course of a run but do not change by large values, so we do not explicitly consider time-dependent calibration uncertainties in our analysis. For further information on O2 calibration techniques, see discussions in \citet{2019ApJ...879...10A}. 

The full raw strain data from the O2 run is publicly available from the Gravitational Wave Open Science Center\footnote{\url{https://www.gw-openscience.org/data}} \citep{2015JPhCS.610a2021V,2019arXiv191211716T}.
For the LIGO O3 data set, the analysis uses the ``C01'' calibration. 
The C01 calibration has estimated maximum amplitude and phase uncertainties of $\sim 7\%$ and $\sim 4 \deg$, respectively \citep{Sun_2020}, which we use as conservative estimates of the true calibration uncertainty near the frequencies analyzed here.
For the Virgo O3 data set, we use the ``V0'' calibration with estimated maximum amplitude and phase uncertainties of 5\% and 2~$\deg$, respectively.

\subsection{Search pipeline}

The time-domain Bayesian method performs a coherent analysis of the interferometers' data, meaning that we analyze the entire data set with an effective single Fourier Transform, thereby preserving the phase information. First, the raw strain data are heterodyned \citep{Dupuis:2005} using the expected signal phase evolution, known precisely from the electromagnetic timing ephemeris. Then a low-pass filter with a knee frequency of 0.25 Hz is applied, and the data are downsampled so that the sampling time is one minute, compared to 60 microseconds originally. This heterodyning is performed for an expected signal whose frequency is at once or twice the rotational frequency of the pulsar.
The heterodyned data is the input to a nested sampling algorithm that is a part of the {\sc
LALInference} package \citep{2010PhRvD..81f2003V, PhysRevD.91.042003}, which infers the unknown signal parameters depending on the model of \gwh emission.

\psrname glitched three times over the course of the \gwh observations (see Figure~\ref{fig:glitch}). For each glitch, we assume an unknown phase offset between the electromagnetic and \gwh phase. The individual phase offsets of multiple glitches that occurred between O2 and O3 cannot be disentangled, so only one phase offset is included for these glitches. This means that we introduce four additional phase parameters when performing parameter estimation. 

We also make use of restricted and unrestricted priors when performing the analysis. In the first case, we use estimates of the orientation of the pulsar relative to the Earth based on a model fit of the observed pulsar wind nebulae torus \citep{Ng:2008}, which imply narrow priors in our analysis on the polarization and inclination angles. Therefore, we use a Gaussian prior on $\psi$ of $2.2864\pm0.0384$\,rad and a bimodal Gaussian prior on $\iota$ with modes at $1.522\pm0.016$ and 
$1.620\pm0.016$\,rad \citep[see][for reasons behind the bimodality]{2015MNRAS.453...53J}. This range of  $\iota$ would suggest the pulsar's rotation axis is almost perpendicular to the line-of-sight, which would in turn lead to a linearly polarized \gwh signal dominated by the `+' polarization component. The second case assumes a uniform isotropic prior on the axis direction, which therefore does not rely on {assumptions about the pulsar's orientation matching that of the wind nebula or uncertainties in the above modeling of the not-well-resolved X-ray observations}. The initial signal phase and glitch phase offsets all use uniform priors over their full ranges. For the single harmonic search, we parameterize the signals using the mass quadrupole $Q_{22}$ and distance. As a conservative approach, we use an unphysical flat prior on $Q_{22}$ with a lower bound at zero and an upper bound of $\scinum{5}{37}$~kg\,m$^2$, which is well above the largest upper limits found in \citet{2019ApJ...879...10A}. For the distance, we use a Gaussian prior with mean of 49.59\,kpc and standard deviation of 0.55\,kpc based on the value given in \citet{pietrzynskietal19}, combining the statistical and systematic errors in quadrature. For the dual harmonic search, which uses the amplitudes $C_{21}$ and $C_{22}$ rather than the physical parameters of $Q_{22}$ and $d$, we use flat priors that are bounded between zero and $\scinum{1}{-22}$, which is again well above the limit implied in \citet{2019ApJ...879...10A}.
To analyze multiple detectors' data sets simultaneously, we combine the product of the {Student's $t$-likelihoods} calculated for each detector \citep{Dupuis:2005}.

The outputs of the analysis are posterior distributions of the parameters of interest, which are $h_0/Q_{22}/\varepsilon$ for the single harmonic search, $C_{21}$ and $C_{22}$ for the dual harmonic search, and the angles $\cos\iota$ and $\psi$ for both choices of priors. In Section~\ref{sec:results}, we present results on the amplitude parameters marginalized over the rest of the parameter space.
{We also provide odds ratios between two hypotheses: the data contain a coherent signal in the detectors, or incoherent signals or noise in the different detectors. These values are used to assess the presence of a signal in the data and, for a given prior choice, can be thought of as a ``detection statistic''.}

\section{Results}\label{sec:results}

Results from our searches do not show evidence for \gwh emission from \psrname via the two models that we assume.
{For the single harmonic model, the Bayesian odds of the data containing a coherent signal between detectors versus incoherent signals or noise in the different detectors \citep[see equation A6 of][]{2017ApJ...839...12A} favor the latter case by $\sim 20\,000$ and $\sim 31\,000$ for the unrestricted and restricted priors, respectively. For the dual harmonic model, the case of an incoherent signal or noise in the detectors is favored by $\lesssim 2\!\times\!10^{8}$ for both prior choices.}

\begin{figure}[htb]
    \centering
    \includegraphics[width=\columnwidth]{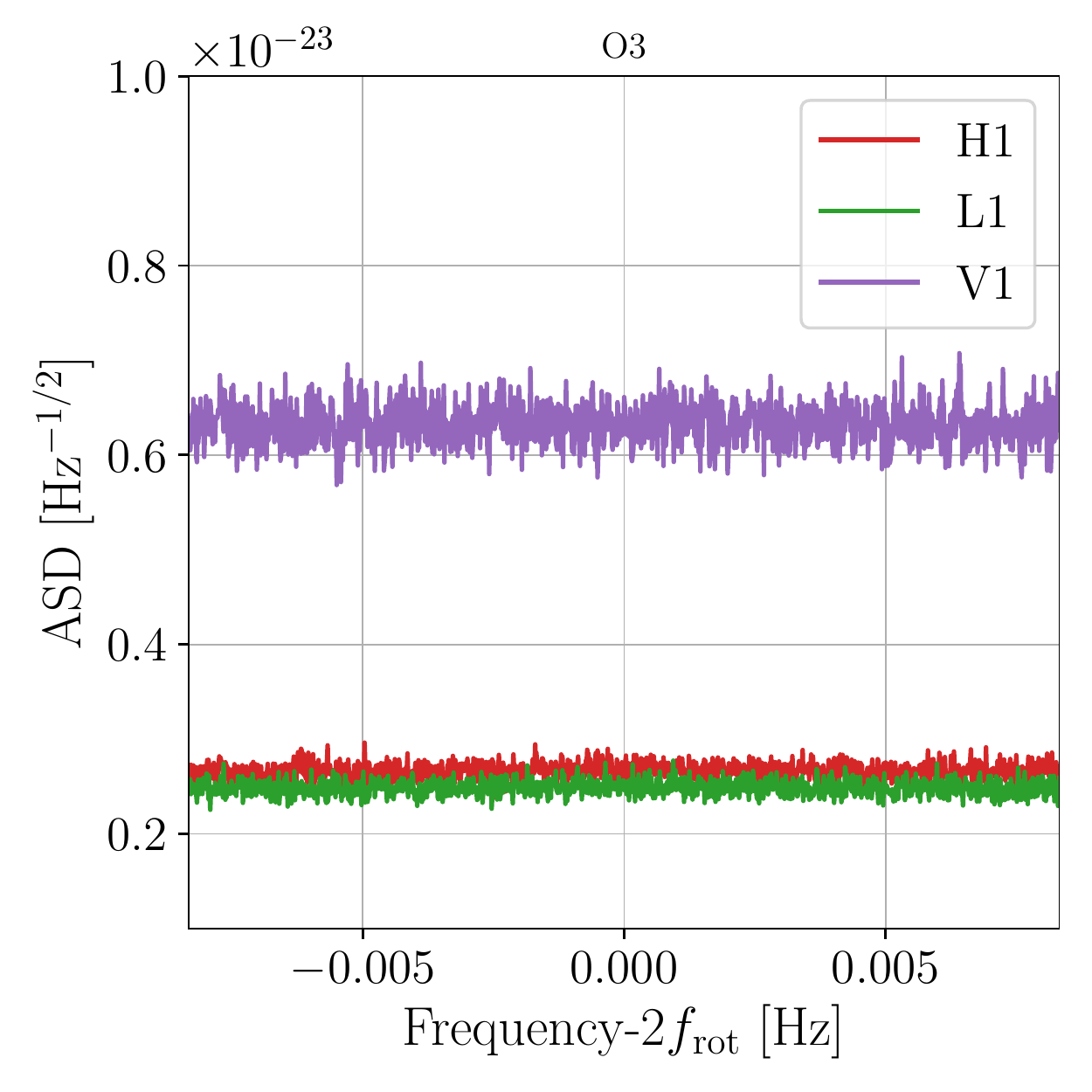}
    \caption{Two-sided amplitude spectral density (ASD) after heterodyning, low pass filtering, and downsampling the raw strain data for the $l=m=2$ \gwh mode. 
    Different color lines indicate the Hanford (H1), Livingston (L1), and Virgo (V1) detectors.}
    \label{fig:amph}
\end{figure}

An amplitude spectral density obtained after the heterodyne correction is displayed in Figure~\ref{fig:amph} for each of the three detectors. If a loud continuous \gwh signal was present, we would expect to see a narrow line feature in the spectrum. The amplitude spectral densities also give an estimation of the sensitivity of the search.

Given the lack of evidence for a signal from either the single or dual harmonic models, we expect the odds between these models to favor the simpler single harmonic model. {Indeed, we find that the single harmonic model is strongly favored by factors of $\sim 5700$ and $\sim 9200$ for the restricted and unrestricted orientation cases, respectively. However, it is worth noting that the odds between models will depend on our choice of the uniform prior range on the amplitude parameters.}

Though no gravitational waves are detected, we can still determine upper limits on  possible \gwh emission from \psrname. Here, we use 95\% credible upper bounds on the amplitude parameters based on their marginalized probability distributions.\footnote{{Simulations on independent and identically distributed noise realizations show that the different noise instantiations can produce upper limits that vary by $\sim 20\%$ at a $1\sigma$ confidence level. However, the Bayesian credible limits we present are valid for our particular dataset.}} The dimensionless \gwh amplitude $h_0$ and coefficients $C_{21}$ and $C_{22}$ are constrained for the single and dual harmonic searches, respectively. For the single harmonic search, $h_0$ can be mapped to a limit on the maximum ellipticity $\varepsilon$ using equation \eqref{eq:h0}. In Table~\ref{table:1}, we show the different constraints for both searches using all O3 data and the last $\sim 6$ days of O2 data (see Section~\ref{sec:gwdata}).
In addition to the detector calibration uncertainties discussed in Section~\ref{sec:gwdata}, we estimate that the statistical uncertainty on the upper
limits due to the use of a finite number of posterior samples is on the order of $1\%$.

\input{table_results.tex}

\begin{figure}[hb]
    \centering
    \includegraphics[width=\columnwidth]{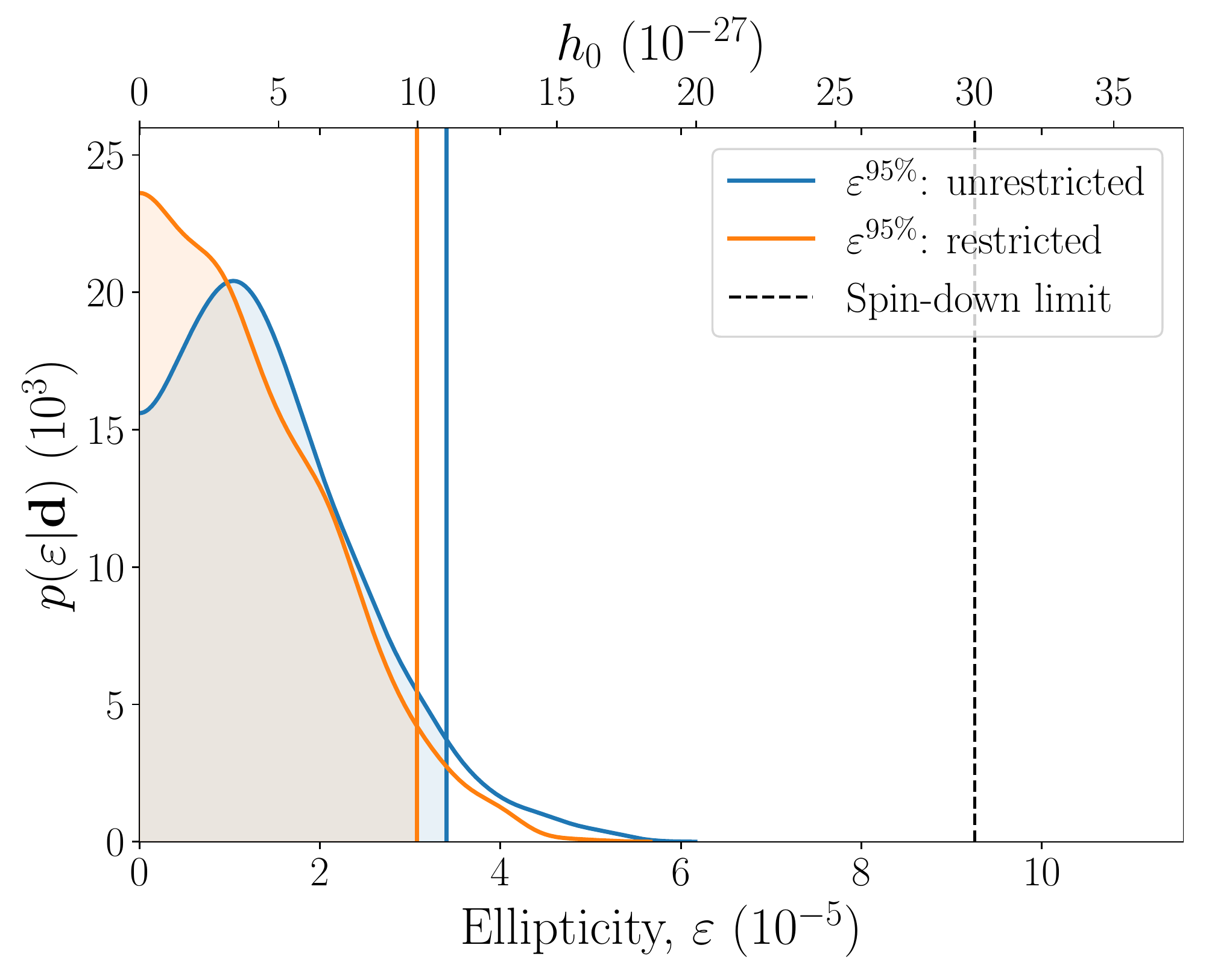}
    \caption{Posterior probability distribution for ellipticity and $h_0$ for the analyses with unrestricted and restricted priors on the pulsar orientation. The 95\% credible upper limits are shown as vertical colored lines, while the spin-down limit is given by the vertical dashed black line.}
    \label{fig:corner}
\end{figure}

Figure~\ref{fig:corner} shows the marginalized posterior probability distributions on the pulsar ellipticity and $h_0$ for the single harmonic search with unrestricted and restricted source orientation priors.
The posteriors show significant support at ellipticities of zero, indicating no evidence of a signal at current sensitivities. We therefore show 95\% credible upper limits on the ellipticity for both prior choices along with the fiducial spin-down limit.

\begin{figure*}[htb]
    \centering
    \includegraphics[width=1.0\textwidth]{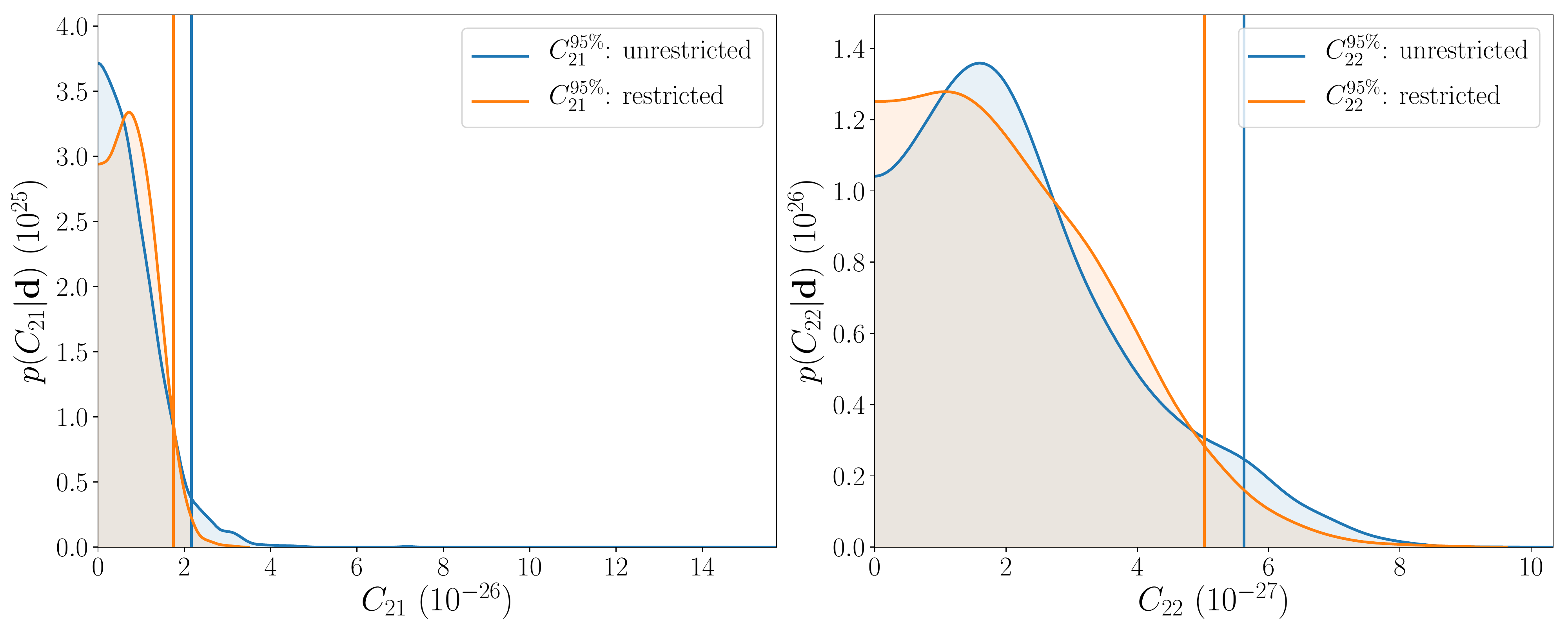}
    \caption{Posterior probability distributions for the amplitudes $C_{21}$ and $C_{22}$ with unrestricted and restricted priors on the pulsar orientation. The 95\% credible upper limits are shown as vertical colored lines.}
    \label{fig:c21c22}
\end{figure*}

Figure \ref{fig:c21c22} shows a similar posterior distribution on the dimensionless amplitudes $C_{21}$ and $C_{22}$ for the dual harmonic model. For this model, no evidence of \gws is found, so an upper limit at 95\% is indicated in both panels of this figure. The model given by equation~(\ref{eq:model}) implies that the value of $C_{21}$ becomes completely unconstrained when $\sin{\iota} = 0$. For the unrestricted orientation prior result shown in the left panel of Figure~\ref{fig:c21c22}, this leads to a long high amplitude tail in the $C_{21}$ posterior distribution. In Figures~\ref{fig:corner} and \ref{fig:c21c22}, we see that the amplitude posteriors can peak away from zero. This behavior was unsurprising and can occur even for pure Gaussian noise. Even with these peaks, the posteriors are still entirely consistent with zero ellipticity. For example, for the unrestricted posterior distribution shown in Figure~\ref{fig:corner}, a value of zero ellipticity is within the minimum 66\% credible interval around the mode.

In contrast to emission in the single harmonic case, an energy-based limit on \gwh emission is rather complex in the dual harmonic case.  The relevant constraint is that the observed spin-down energy is equal to the sum of the luminosities at the two harmonics. These two emissions have different beam patterns: the emission at the rotation frequency is strongest along the rotational equator ($\iota = \pi/2$ direction), where the polarization is linear, while emission at twice the rotation frequency is strongest along the axis of rotation ($\iota = 0$), where the polarization is circular.  Therefore, the spin-down limit on the maximum amplitudes of the two harmonics  depends on both the relative size of the intrinsic strength of the two components and the orientation of the spin axis relative to the observer. To provide some insight, if we compare the sky-averaged emission strength at only the rotation frequency to emission at only twice the rotation frequency, the spin-down limit would allow the amplitude of the radiation at the rotation frequency to be approximately twice as strong as that at twice the rotation frequency \citep[see Section~3 of][for more details]{2010MNRAS.402.2503J}.

The results presented above use all O3 data in combination with about six days of O2 data, when \NICER was operating and monitoring \psrname.
We also conducted searches using only O3 data or using O3 data plus O2 data from 2017 March 22 to the end of O2.  The latter analysis assumes no glitches occurred during the additional time and represents the estimated maximum time that can be safely included without a contemporaneous timing model (see Section~\ref{sec:gwdata}).
For only O3 data, we obtain $h_0$ and $\varepsilon$ limits that are worse by $\sim 7\%$ for the unrestricted prior and unchanged for the restricted prior, compared to those from those shown in Table~\ref{table:1}.
For O3 data plus the extra O2 data, we obtain amplitude limits that are improved by $\lesssim 20\%$ compared to those shown in Table~\ref{table:1}.

\section{Conclusions}

Using data from LIGO/Virgo's second and third observing runs, we searched for mass quadrupolar-sourced \gws from the young, dynamic \psrname
at once or twice the pulsar's rotational frequency of 62\,Hz. For the first time, we reached below the gravitational-wave spin-down limit for \psrname and showed that \gwh emission for a pure $l=m=2$ mode accounts for less than 14\% of the pulsar's  spin-down energy budget.  We placed the third most stringent constraint on the ellipticity ($\varepsilon<3\times 10^{-5}$) of any young pulsar (behind only the Crab pulsar and B1951+32/J1952+3252; \citealt{2019ApJ...879...10A,abbott2020gravitational}).
While this limit is much higher than those of old recycled millisecond pulsars (for which  $\varepsilon<10^{-8}$; \citealt{abbott2020gravitational}), young pulsars such as \psrname and the Crab pulsar are important because they have much stronger magnetic fields (and are hotter) and thus might have greater ellipticities.  The ellipticity constraint of \psrname is also {above or} near estimates of the maximum ellipticity that can be sustained by an elastically deformed neutron star crust \citep{2013PhRvD..88d4004J,2018PhRvL.121m2701C,gittins2021}.

\psrname is a frequently glitching pulsar and potential source of continuous \gws. The X-ray data from \NICER gives us the necessary tools to account for the phase evolution of a \gwh signal over time, which allows us to perform a fully coherent and sensitive search for such a signal. While our multi-messenger analysis focuses on \gws from a time-varying mass quadrupole ($n=5$), another search could be performed for \gws from a r-mode fluid oscillation ($n=7$) using wider-band techniques (e.g., \citealt{fesikpapa20,fesikpapa20b}, using O2 data).
The strain sensitivity achieved in our analysis ($1\times 10^{-26}$) is also comparable to the $(2-3)\times10^{-26}$ estimated in \cite{anderssonetal18} for r-mode emission from \psrname.

Finally, from the observed correlation between glitch size and time-to-next-glitch
for \psrname
\citep{2006ApJ...652.1531M,antonopoulouetal18,2018ApJ...852..123F,hoetal20},
we can hope to measure in the future low braking indices (7 or even lower)
after the largest glitches.
As noted above, braking indices of 5 and 7 are predicted by \gw-emitting mechanisms.
The observed evolution of $\nig$ to lower values than those shown in Figure~\ref{fig:tnig}, which may occur after the effects of glitches on the pulsar's spin-down behavior have decayed, may indicate that \gws are continuously emitted between glitches.
On the other hand, glitches may trigger detectable transient \gws
\citep{prixetal11,hoetal20b,yimjones20}, and \gwh searches at glitch epochs of other pulsars have been conducted \citep{keiteletal19}.
It is therefore vital to continue to monitor the spin evolution of \psrname,
not only to obtain the timing ephemeris and measure braking indices,
but also to know when this pulsar undergoes a glitch.
Since the spin period of \psrname is only detectable at X-ray energies,
\NICER is the only effective means to perform the necessary observations.
Fortunately \NICER is anticipated to operate until at least late 2022,
overlapping with the fourth observing run of LIGO/Virgo and KAGRA \citep{KAGRA},
which is likely to begin in 2022 and continue into 2023.

\section*{Acknowledgments}
\input{acknowledgements}

D.An.\ acknowledges support from an EPSRC fellowship (EP/T017325/1). C.M.E.\ acknowledges support from FONDECYT/Regular 1171421 and USA1899-Vridei 041931SSSA-PAP (Universidad de Santiago de Chile, USACH). W.C.G.H.\ acknowledges support through grants 80NSSC19K1444 and 80NSSC21K0091 from NASA.  This work is supported by NASA through the \NICER mission and the Astrophysics Explorers Program and uses data and software provided by the High Energy Astrophysics Science Archive Research Center (HEASARC), which is a service of the Astrophysics Science Division at NASA/GSFC and High Energy Astrophysics Division of the Smithsonian Astrophysical Observatory.
\facility{\NICER}

\bibliographystyle{aasjournal}
\bibliography{references} 

\iftoggle{endauthorlist}{
  \let\author\myauthor
  \let\affiliation\myaffiliation
  \let\maketitle\mymaketitle
  \title{The LIGO Scientific Collaboration, Virgo Collaboration, and KAGRA Collaboration  }
  \pacs{}

\input{KAGRA-LSC-Virgo-Authors-Aug-2020-aas.tex}\input{authorlist_nicer.tex}  \newpage
  \maketitle
}

\label{lastpage}
\end{document}

%% file: KAGRA-LSC-Virgo-Authors-Aug-2020-aas.tex
\AuthorCollaborationLimit=3000  


\author{R.~Abbott}
\affiliation{LIGO, California Institute of Technology, Pasadena, CA 91125, USA}
\author{T.~D.~Abbott}
\affiliation{Louisiana State University, Baton Rouge, LA 70803, USA}
\author{S.~Abraham}
\affiliation{Inter-University Centre for Astronomy and Astrophysics, Pune 411007, India}
\author{F.~Acernese}
\affiliation{Dipartimento di Farmacia, Universit\`a di Salerno, I-84084 Fisciano, Salerno, Italy  }
\affiliation{INFN, Sezione di Napoli, Complesso Universitario di Monte S.Angelo, I-80126 Napoli, Italy  }
\author{K.~Ackley}
\affiliation{OzGrav, School of Physics \& Astronomy, Monash University, Clayton 3800, Victoria, Australia}
\author{A.~Adams}
\affiliation{Christopher Newport University, Newport News, VA 23606, USA}
\author{C.~Adams}
\affiliation{LIGO Livingston Observatory, Livingston, LA 70754, USA}
\author{R.~X.~Adhikari}
\affiliation{LIGO, California Institute of Technology, Pasadena, CA 91125, USA}
\author{V.~B.~Adya}
\affiliation{OzGrav, Australian National University, Canberra, Australian Capital Territory 0200, Australia}
\author{C.~Affeldt}
\affiliation{Max Planck Institute for Gravitational Physics (Albert Einstein Institute), D-30167 Hannover, Germany}
\affiliation{Leibniz Universit\"at Hannover, D-30167 Hannover, Germany}
% ...

\collaboration{The LIGO Scientific Collaboration, the Virgo Collaboration and the KAGRA Collaboration}



%% file: authorlist_nicer.tex
\author{D. Antonopoulou}
  \affiliation{Nicolaus Copernicus Astronomical Center, Polish Academy of Sciences, ul.\ Bartycka 18, 00-716 Warsaw, Poland}
  \affiliation{Jodrell Bank Centre for Astrophysics, School of Physics and Astronomy, University of Manchester, Manchester M13 9PL, United Kingdom}
  \author{Z. Arzoumanian}
  \affiliation{X-Ray Astrophysics Laboratory, NASA Goddard Space Flight Center, Greenbelt, MD 20771, USA}
  \author{T. Enoto}
  \affiliation{Extreme Natural Phenomena RIKEN Hakubi Research Team, RIKEN Cluster for Pioneering Research, 2-1 Hirasawa, Wako, Saitama 351-0198, Japan}
  \author{C.~M. Espinoza}
  \affiliation{Departamento de F\'isica, Universidad de Santiago de Chile, Avenida Ecuador 3493, 9170124 Estaci\'on Central, Santiago, Chile}
  \author{S. Guillot}
  \affiliation{IRAP, CNRS, 9 avenue du Colonel Roche, BP 44346, F-31028 Toulouse Cedex 4, France}
  \affiliation{Universit\'e de Toulouse, CNES, UPS-OMP, F-31028 Toulouse, France}
  

%% file: table_results.tex
\begin{deluxetable}{cccccc}[htb]
\tablecaption{95\% upper limits on \gwh strain, ellipticity, and other quantities based on unrestricted (UR) and restricted (R) choices for priors on polarization and inclination angles. Results here come from analyzing all O3 data and the last 6 days of O2 data.
\label{table:1}
}
\tablewidth{0pt}
\tablehead{
 \colhead{Prior} 
 & \colhead{$h_0^{95\%}$} 
 & \colhead{$\varepsilon^{95\%}$} 
 & \colhead{$h_0^{95\%}/h_0^{\text{sd}}$}
  & \colhead{$C_{21}^{95\%}$} 
 & \colhead{$C_{22}^{95\%}$} 
 \\  
 ~ 
 & \colhead{($10^{-26}$)} 
 & \colhead{($10^{-5}$)}
 & ~ 
  & \colhead{($10^{-26}$)} 
 & \colhead{($10^{-27}$)}
 }
 \startdata
 UR 
 & 1.1
 & 3.4
 & 0.37
  & 2.2
 & 5.6 \\ 
 R 
 & 1.0
 & 3.1 
 & 0.33
  & 1.8
 & 5.0
\enddata
\end{deluxetable}

%% file: acknowledgements.tex
The authors gratefully acknowledge the support of the United States
National Science Foundation (NSF) for the construction and operation of the
LIGO Laboratory and Advanced LIGO as well as the Science and Technology Facilities Council (STFC) of the
United Kingdom, the Max-Planck-Society (MPS), and the State of
Niedersachsen/Germany for support of the construction of Advanced LIGO
and construction and operation of the GEO600 detector.
Additional support for Advanced LIGO was provided by the Australian Research Council.
The authors gratefully acknowledge the Italian Istituto Nazionale di Fisica Nucleare (INFN),
the French Centre National de la Recherche Scientifique (CNRS) and
the Netherlands Organization for Scientific Research,
for the construction and operation of the Virgo detector
and the creation and support  of the EGO consortium.
The authors also gratefully acknowledge research support from these agencies as well as by
the Council of Scientific and Industrial Research of India,
the Department of Science and Technology, India,
the Science \& Engineering Research Board (SERB), India,
the Ministry of Human Resource Development, India,
the Spanish Agencia Estatal de Investigaci\'on,
the Vicepresid\`encia i Conselleria d'Innovaci\'o, Recerca i Turisme and the Conselleria d'Educaci\'o i Universitat del Govern de les Illes Balears,
the Conselleria d'Innovaci\'o, Universitats, Ci\`encia i Societat Digital de la Generalitat Valenciana and
the CERCA Programme Generalitat de Catalunya, Spain,
the National Science Centre of Poland,
the Swiss National Science Foundation (SNSF),
the Russian Foundation for Basic Research,
the Russian Science Foundation,
the European Commission,
the European Regional Development Funds (ERDF),
the Royal Society,
the Scottish Funding Council,
the Scottish Universities Physics Alliance,
the Hungarian Scientific Research Fund (OTKA),
the French Lyon Institute of Origins (LIO),
the Belgian Fonds de la Recherche Scientifique (FRS-FNRS),
Actions de Recherche Concertées (ARC) and
Fonds Wetenschappelijk Onderzoek – Vlaanderen (FWO), Belgium,
the Paris \^{I}le-de-France Region,
the National Research, Development and Innovation Office Hungary (NKFIH),
the National Research Foundation of Korea,
Industry Canada and the Province of Ontario through the Ministry of Economic Development and Innovation,
the Natural Science and Engineering Research Council Canada,
the Canadian Institute for Advanced Research,
the Brazilian Ministry of Science, Technology, Innovations, and Communications,
the International Center for Theoretical Physics South American Institute for Fundamental Research (ICTP-SAIFR),
the Research Grants Council of Hong Kong,
the National Natural Science Foundation of China (NSFC),
the Leverhulme Trust,
the Research Corporation,
the Ministry of Science and Technology (MOST), Taiwan
and
the Kavli Foundation.
The authors gratefully acknowledge the support of the NSF, STFC, INFN and CNRS for provision of computational resources. 

This work was supported by MEXT, JSPS Leading-edge Research Infrastructure Program, JSPS Grant-in-Aid for Specially Promoted Research 26000005, JSPS Grant-in-Aid for Scientific Research on Innovative Areas 2905: JP17H06358, JP17H06361 and JP17H06364, JSPS Core-to-Core Program A. Advanced Research Networks, JSPS Grant-in-Aid for Scientific Research (S) 17H06133, the joint research program of the Institute for Cosmic Ray Research, University of Tokyo, National Research Foundation (NRF) and Computing Infrastructure Project of KISTI-GSDC in Korea, Academia Sinica (AS), AS Grid Center (ASGC) and the Ministry of Science and Technology (MoST) in Taiwan under grants including AS-CDA-105-M06, Advanced Technology Center (ATC) of NAOJ, and Mechanical Engineering Center of KEK. 

We thank all essential workers who put their health at risk during the COVID-19 pandemic, without whom we would not have been able to complete this work.